\documentclass[english,a4paper,pre,twocolumn,amsmath,amssymb,superscriptaddress]{revtex4-1}
\usepackage{preamble}

\begin{document}

\title{Many-body correlations from integral geometry}
\date{\today}

\author{Joshua F. Robinson}
\email{joshua.robinson@bristol.ac.uk}
\affiliation{H.\ H.\ Wills Physics Laboratory, University of Bristol, Bristol BS8 1TL, UK}
\author{Francesco Turci}
\affiliation{H.\ H.\ Wills Physics Laboratory, University of Bristol, Bristol BS8 1TL, UK}
\author{Roland Roth}
\affiliation{Institut f\"ur Theoretische Physik, Universit\"at T\"ubingen, 72076 T\"ubingen, Germany}
\author{C. Patrick Royall}
\affiliation{H.\ H.\ Wills Physics Laboratory, University of Bristol, Bristol BS8 1TL, UK}
\affiliation{School of Chemistry, Cantock’s Close, University of Bristol, Bristol, BS8 1TS, UK}
\affiliation{Centre for Nanoscience and Quantum Information, University of Bristol, Bristol BS8 1FD, UK}

\begin{abstract}
  In a recent letter we presented a framework for predicting the concentrations of many-particle local structures inside the bulk liquid as a route to assessing changes in the liquid approaching dynamical arrest.
  Central to this framework was the morphometric approach, a synthesis of integral geometry and liquid state theory, which has traditionally been derived from fundamental measure theory.
  We present the morphometric approach in a new context as a generalisation of scaled particle theory, and derive several morphometric theories for hard spheres of fundamental and practical interest.
  Our central result is a new theory which is particularly suited to the treatment of many-body correlation functions in the hard sphere liquid, which we demonstrate by numerical tests against simulation.
\end{abstract}


\maketitle

\section{Introduction}

Since the beginnings of modern liquid state theory \cite{KirkwoodJCP1935}, the hard sphere liquid has remained the archetypal model for atomic systems and soft matter.
The dynamics of the system at high density in the metastable regime above the freezing transition are hotly debated, despite relentless study.
Proposed mechanisms for dynamical phenomena all loosely fall under the broad umbrella of many-body correlations; nucleation occurs via crystal seed formation \cite{SearJPCM2007}, and to explain dynamic arrest approaching the glass transition thermodynamic theories invoke cooperatively rearranging regions \cite{LubchenkoARPC2007} or elastic soft modes \cite{BritoJCP2009} while kinetic theories posit the existence of dynamical defects \cite{ChandlerARPC2010}.
In a recent letter \cite{RobinsonPRL2019} we proposed a framework for treating many-body correlations, and developed an operational scheme for predicting the populations and dynamics of local structural motifs within a uniform liquid.
Central to this is the use of the \emph{morphometric approach}.

The morphometric approach provides an efficient means of treating the thermodynamics of a bulk liquid without fully determining its equilibrium density profile \cite{KonigPRL2004,RothPRL2006,Hansen-GoosPRL2007,RobinsonPRL2019}.
Detailed investigations have shown that it is highly accurate in the hard sphere liquid regime \cite{OettelEL2009,AshtonPRE2011,LairdPRE2012,BlokhuisPRE2013,UrrutiaPRE2014,Hansen-GoosJCP2014}, so we can expect an accurate treatment where the bulk system provides background depletion interactions while its detailed microstates remain unimportant.
This feature makes it ideally suited for many-body correlations if we can identify relevant dynamical degrees of freedom.
While existing morphometric theories have been proven accurate in the liquid regime, we require a theory which works in the supercooled regime.
Here we derive such a theory using scaled particle theory (SPT).

SPT determines bulk properties from consideration of a spherical solute of varying radius.
It remains one of most enduring theories of simple liquids; though 60 years old as of this year \cite{ReissJCP1959}, aspects of this approach remain in modern theories.
This is particularly true for hard spheres where SPT has been unified with the Percus-Yevick integral equation solution \cite{WertheimPRL1963}, another old theory, in the form of fundamental measure theory (FMT) \cite{RosenfeldPRL1989}.
Though originally a theory of single-component hard spheres \cite{ReissJCP1959}, SPT has been extended to other potentials \cite{ReissJCP1960,HelfandJCP1960,ReissJCP1961} and shapes \cite{GibbonsMP1969,GibbonsMP1970}, mixtures \cite{LebowitzJCP1965}, dimers \cite{StillingerJCP2006,ChatterjeeJCP2006} and disks \cite{HelfandJCP1961,MartinJCP2018,Hansen-GoosJCP2019}.
Morphological thermodynamics can be seen as a modern generalisation of SPT for a wide class of physically relevant geometries.
Its basis in integral geometry replaces the semi-empirical approach of classical SPT with clearly defined postulates.
In this work we present the morphometric approach in the context of SPT and derive a new theory suitable for high densities above freezing.
In additional appendices, we show that minor modifications of our arguments can be used to derive previous theories: the classical SPT coefficients, and the White Bear II morphometric coefficients of Ref.\ \cite{Hansen-GoosJPCM2006}.

In section \ref{sec:many-body-correlations} we show how one can map the problem of treating many-body correlations onto a solvent-solute problem.
We spend the rest of the paper discussing the solvation problem through the lens of SPT.
We introduce the morphometric approach as a useful generalisation of SPT in section~\ref{sec:morphometric-approach}, and derive a theory well-suited for treating many-body correlations using scaled particle arguments.
In section \ref{sec:numerics} we numerically test these theories' two-- and three--body correlation functions to demonstrate their effectiveness in treating correlation functions.

\section{Solvation expression for many-body correlations}
\label{sec:many-body-correlations}

\subsection{Correlations in terms of the insertion cost}

We will show that correlations of $n$ particles at positions $\vec{r}^n := \{\vec{r}_1, \cdots, \vec{r}_n\}$ can be expressed in terms of the free energy cost of inserting them at $\vec{r}^n$, by generalising the \emph{potential distribution theorem} \cite{WidomJCP1963,WidomJPC1982} to many particles.
The classical approach, also known as \emph{Widom's insertion method}, expresses the (excess) chemical potential $\mu^\mathrm{ex}$ of a single-component system as the free energy cost of inserting an additional particle.
See Ref.\ \cite{Rowlinson2002} and references therein for a detailed review of this classical approach.
Our generalisation results in a \emph{potential of mean force} for interactions between the $n$ particles, which is formally identical to the chemical potential of a solute; this latter form is particularly suitable for geometric approximation schemes.

We consider a bulk liquid (the solvent) of $N$ particles with interaction potential energy $U_N$.
Integrating over all solvent arrangements in the absence of any external field gives the (grand-canonical) average
\begin{equation*}
  \left< \cdots \right> =
  \frac{1}{\Xi} \sum_{N=0}^\infty \frac{z^N}{N!} \int \left(\cdots\right) e^{-\beta U_N} \, d\vec{r}^N,
\end{equation*}
with partition function $\Xi := e^{-\beta \Omega_\mathrm{hom}}$, where $\Omega_\mathrm{hom} = -p V$ is the usual homogeneous grand potential.
The activity is $z = \exp{\beta\mu} / \Lambda^d$ in terms of the (total) chemical potential $\mu$ and the thermal de Broglie wavelength $\Lambda$.

Descriptions of many-body correlations naturally employ the $n$-particle density $\rho^{(n)}$, defined as
\begin{equation}\label{eq:n-density-pdf}
  \mathrm{Prob}\left[ \textit{any } n \textrm{ particles in volume } d\vec{r}^n \right]
  :=
  \rho^{(n)}(\vec{r}^n) \, d\vec{r}^n.
\end{equation}
The $n$-density can be obtained by integrating the full (configurational) probability distribution over the remaining degrees of freedom.
For the single-component system this yields~\cite{Hansen2013}
\begin{equation*}
  \rho^{(n)}(\vec{r}^n)
  =
  \frac{1}{\Xi} \sum_{N=n}^\infty \frac{z^N}{(N-n)!} \int e^{-\beta U_N} \, d\vec{r}^{(N-n)}.
\end{equation*}
Changing the summation limits $N \rightarrow N+n$ we obtain
\begin{equation}\label{eq:n-density}
\begin{aligned}
  \rho^{(n)}(\vec{r}^n)
  &=
  \frac{z^n}{\Xi} \sum_{N=0}^\infty \frac{z^N}{N!} \int e^{-\beta U_{N+n}} \, d\vec{r}^{N}
  \\ &=
  z^n e^{-\beta U_n} \left< e^{-\beta U_{n \leftrightarrow N}} \right>
\end{aligned}
\end{equation}
where in the latter step we decomposed the total potential $U_{N+n}$ into purely local and solvent terms, i.e.\ $U_{N+n} = U_n + U_N + U_{n \leftrightarrow N}$, where $U_\alpha$ for $\alpha \in \{n,N\}$ indicates the internal interactions between particles in component $\alpha$.
The ``interspecies'' interactions are contained within $U_{n \leftrightarrow N}$ which acts as an external field for the solvent.
Thus, \eqref{eq:n-density} becomes
\begin{equation*}
  \rho^{(n)}(\vec{r}^n)
  =
  z^n e^{-\beta (U_n + \Omega - \Omega_\mathrm{hom})}.
\end{equation*}
where $\Omega$ is the grand potential of the solvent in the presence of the $n$-particle inhomogeneity.
Splitting the chemical potential into its ideal and excess parts so that $\beta\mu = \ln{\Lambda^d \rho} + \beta\mu^\mathrm{ex}$ gives
\begin{equation*}
  \rho^{(n)}(\vec{r}^n)
  =
  \rho^n e^{-\beta (U_n + \Omega - \Omega_\mathrm{hom} - n\mu^\mathrm{ex})}.
\end{equation*}
The $n$-particle distribution functions are then determined from~\cite{Hansen2013}
\begin{equation}\label{eq:distribution-functions}
  g^{(n)}(\vec{r}^n)
  := \frac{\rho^{(n)}(\vec{r}^n)}{\rho^n}
  = e^{-\beta(U_n + \Delta\Omega - n\mu^\mathrm{ex})}
\end{equation}
where $\Delta\Omega := \Omega - \Omega_\mathrm{hom}$ is the reversible (free energy) cost of inserting the particles at fixed position $\vec{r}^n$, or equivalently describes the average depletion interactions between mobile particles.
For $n=1$ we have $\Delta\Omega = \mu^\mathrm{ex}$ and this is identical to the potential distribution theorem of Widom \cite{WidomJCP1963,WidomJPC1982}.
The distribution functions are written in terms of the potential
\begin{equation}\label{eq:potential-mean-force}
  \begin{split}
    \phi^{(n)}(\vec{r}^n) &:=
    - k_B T \ln{g^{(n)}(\vec{r}^n)} \\
    &=
    U_n + \Delta\Omega - n\mu^\mathrm{ex},
  \end{split}
\end{equation}
which we call the \emph{generalised potential of mean force}.
For the case $n=2$ this reduces to the usual potential of mean force in the liquid state literature \cite{Hansen2013}.

This completes our proof that the correlations can be transformed to a potential, and we can proceed with a geometrical construction for $\Delta \Omega$.

\begin{figure}
  \includegraphics[width=\linewidth]{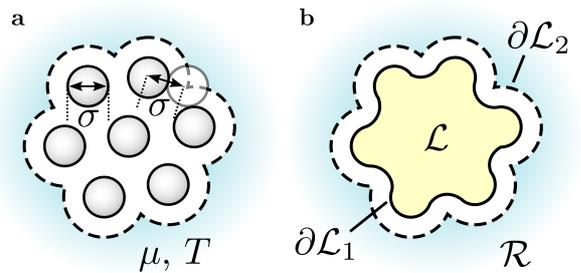}
  \caption{
    The system considered for many-body correlations showing
    (a) the local particles surrounded by the remaining liquid acting as a thermal reservoir at fixed chemical potential and temperature, and
    (b) possible partitions of space into the local $\mathcal{L}$ and remaining $\mathcal{R}$ components for two choices of dividing surface: $\partial\mathcal{L}_1$ is the molecular surface while $\partial\mathcal{L}_2$ is the solvent accessible surface (see discussion around Eq.\ \eqref{eq:exclusion-transform}).
  }
  \label{fig:system}
\end{figure}

\subsection{Representing the insertion cost as a solvation problem}

For systems with excluded volume interactions, we can divide the space into a local component $\mathcal{L} \subset \mathbb{R}^d$ of volume $V_\mathcal{L}$ inaccessible to solvent degrees of freedom and the remaining space $\mathcal{R} = \mathbb{R}^d \setminus \mathcal{L}$ of volume $V_\mathcal{R}$ filled by the rest of the liquid (Fig. \ref{fig:system}).
The total volume is $V = V_\mathcal{L} + V_\mathcal{R}$ so the homogeneous grand potential is
\begin{equation*}
  \Omega_\mathrm{hom} = -p V.
\end{equation*}
After inserting the inhomogeneity the total volume accessible to the rest of the liquid will be reduced by $V_\mathcal{L}$, so the grand potential becomes
\begin{equation*}
  \Omega = -p V_\mathcal{R} + \Omega_\mathrm{ex}[\partial\mathcal{L}],
\end{equation*}
where $\Omega_\mathrm{ex}$ is an excess term brought about by the introduction of a dividing surface $\partial\mathcal{L}$ between the two liquid components.
Subtracting these two expressions gives
\begin{equation*}
  \Delta \Omega
  := \Omega - \Omega_\mathrm{hom}
  = p V_\mathcal{L} + \Omega_\mathrm{ex}[\partial\mathcal{L}].
\end{equation*}
This dividing surface has area $A_{\partial\mathcal{L}}$, creating a surface tension $\gamma$ so we can write the excess term as
\begin{equation*}
  \Omega_\mathrm{ex}[\partial\mathcal{L}] =
  \gamma[\partial\mathcal{L}] A_{\partial\mathcal{L}}
\end{equation*}
which is a formal definition of surface tension and depends on the choice of dividing surface (see two examples in Fig.\ \ref{fig:system}b).
We know from density functional theory \cite{EvansAP1979} that the excess free energy is a functional of the density profile, which will in turn depend on the shape of the boundary; we write $\gamma = \gamma[\partial \mathcal{L}]$ to indicate this functional dependence on the surface shape.
The \emph{solvation form} of the inhomogeneous grand potential term in \eqref{eq:potential-mean-force} is then
\begin{equation}\label{eq:surface-tension}
  \Delta \Omega[\mathcal{L}] =
  p V_\mathcal{L} + \gamma[{\partial\mathcal{L}}] A_{\partial\mathcal{L}}.
\end{equation}
The problem of determining the $n$-particle distributions has been reduced to a solvation problem: we must find the surface tension between a solute (the specific local arrangement) and a solvent (the rest of the liquid).
We will use the solute--solvent terminology, but one could also think of local--bulk nomenclature.

\section{Obtaining a morphological theory for many-body correlations}
\label{sec:morphometric-approach}

We will consider a single-component hard sphere fluid, for particles of diameter $\sigma$ and bulk volume fraction $\eta$.
Using the correspondence between many-body correlation functions and chemical potentials, we require an approximate model for solvation and a choice of surface in \eqref{eq:surface-tension} to evaluate $\Delta \Omega$ in \eqref{eq:potential-mean-force}.
We introduce our central approximation in section \ref{sec:morph-ansatz} and our choice of surface in \ref{sec:surface}.
Then, we show that previous theories fail to produce accurate correlation functions at high densities in \ref{sec:spt-failure} and derive a new theory to rectify this in \ref{sec:virial-spt}.

\subsection{Our central approximation: the morphometric/scaled particle ansatz}
\label{sec:morph-ansatz}

Our key approximation, the morphometric approach, can be understood as a generalisation of scaled particle theory.
In every formulation of scaled particle theory one considers a hard \emph{spherical} solute of radius $R$.
In most approaches, the cost $\Delta \Omega$ is assumed to have an analytic expansion in powers of the radius; in classical approaches this was simply postulated, however we will be able provide proper justification below through geometric arguments.
Recognising that terms scaling faster than $R^3$ must be zero for it to remain well-defined in the limit of large solutes leads to the third-order polynomial \cite{ReissJCP1959}
\begin{equation}\label{eq:spt-ansatz}
  \Delta\Omega(R) =
  p \, \frac{4\pi R^3}{3} + a_2 \, 4 \pi R^2 + a_1 \, 4 \pi R + a_0 \, 4 \pi,
\end{equation}
where we identified the largest power with the work term $pV$ from comparison with \eqref{eq:surface-tension}, and $\{a_0, a_1, a_2\}$ are thermodynamic coefficients describing the subleading corrections.
We have chosen to introduce factors of $4\pi$ in front of the subleading terms to lead into the generalisation beyond spherical geometries.
For a general solute $K \subset \mathbb{R}^3$ we then write the morphometric insertion cost as
\begin{equation}\label{eq:morph-ansatz}
  \Delta\Omega[K] =
  p V[K]
  + a_2 A[K]
  + a_1 C[K]
  + a_0 X[K],
\end{equation}
where $C$ and $X$ are the integrated mean and Gaussian curvatures.
All of these functionals act on $K$ but the latter three can also be \emph{expressed} as surface integrals, as in
\begin{subequations}
  \begin{align}
    A[K]
    &=
    \int_{\partial K} \, dA
    \\
    C[K]
    &=
    \frac{1}{2} \int_{\partial K} \Tr{\kappa} \, dA
    \\
    X[K]
    &=
    \int_{\partial K} \det{\kappa} \, dA
  \end{align}
\end{subequations}
where $\kappa$ is the curvature tensor for the surface $\partial K$.
For a spherical solute these reduce to the values given in \eqref{eq:spt-ansatz}, so this represents a proper generalisation of SPT for more general geometries.
We give a brief justification of the \emph{ansatz} \eqref{eq:morph-ansatz} using integral geometric arguments in appendix \ref{appendix:spt-ansatz}.

The key advantage of a geometric expansion of the free energy is that the role of thermodynamics and geometry are kept separate.
Thermodynamics only enters through the coefficients $\{p,a_2,a_1,a_0\}$, so they can be determined in simple geometries to obtain a general theory.
As a linear theory, \emph{only} four (independent) equations are required to fix these coefficients; with many thermodynamic relations to choose from this approximate theory is overconstrained in general.
We must use physical intuition to choose suitable equations, after which the accuracy of the resulting coefficients can be assessed.
After determining these coefficients all the complexity of computing $\Delta \Omega$ is reduced to measuring the geometric quantities $\{V,A,C,X\}$ of the specific solute.
However, we must first specify a choice of surface $\partial \mathcal{L}$ in \eqref{eq:surface-tension} before we can proceed.

\begin{figure}
 \includegraphics[width=\linewidth]{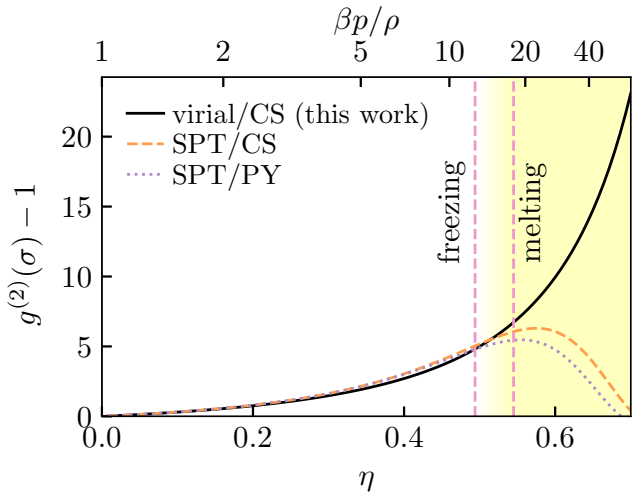}
 \caption{
   Contact values of the radial distribution function against volume fraction $\eta$ and reduced pressure for the hard sphere liquid with \eqref{eq:distribution-functions} and \eqref{eq:morph-ansatz} for the explicit form of $g^{(2)}$, assuming the Carnahan-Starling (CS) and scaled particle theory/Percus-Yevick (SPT/PY) equations of state.
   Contact values are determined with three sets of morphometric coefficients: virial/CS, derived in this work to be quasi-exact (i.e.\ satisfying the virial theorem \eqref{eq:contact-g}) by construction; SPT/CS, a generalisation of scaled particle theory which imposes the CS equation of state; and SPT/PY, the classical scaled particle solution.
   The latter two scaled particle theories feature a spurious decay in the supercooled regime (shaded area).
   The hard sphere freezing and melting volume fractions are indicated by pink dashed lines to show the onset of the supercooled regime.
 }
\label{fig:contact-g}
\end{figure}

\subsection{Choice of dividing surface}
\label{sec:surface}

All coefficients we give are for the molecular geometry bounded by the \emph{molecular surface} ($\partial \mathcal{L}_1$ in Fig.\ \ref{fig:system}b), the surface where interactions occur between the solute and a test particle representing the remaining liquid.
However, it is usually more convenient to do calculations with the excluded geometry: the space inaccessible to the \emph{centre} of a test particle bounded by the \emph{solvent accessible surface} ($\partial \mathcal{L}_2$ in Fig.\ \ref{fig:system}b).
Note that there is also an infinite family of well-defined parallel surfaces between these two extremes, but they are not widely used in practice so we will not consider them \cite{OettelEL2009}.
The choice of dividing surface will change the surface tension, and thus requires new coefficients $\{a_0', a_1', a_2'\}$ i.e.\
\begin{equation}
  \Delta\Omega[K]
  =
  p V_+[K]
  + a_2' A_+[K]
  + a_1' C_+[K]
  + a_0' X_+[K],
\end{equation}
where the excluded geometry terms transform via the canonical relations \cite{Hansen-GoosJPCM2006,OettelEL2009,Santalo2004,Klain1997}
\begin{subequations}\label{eq:exclusion-transform-volumes}
  \begin{align}
    X_+[K]
    &=
    X[K],
    \\
    C_+[K]
    &=
    C[K] + \frac{\sigma}{2} X[K],
    \\
    A_+[K]
    &=
    A[K] + \sigma \, C[K] + \frac{\sigma^2}{4} X[K],
    \\
    V_+[K]
    &=
    V[K]
    + \frac{\sigma}{2} A[K]
    + \frac{\sigma^2}{4} C[K]
    + \frac{\sigma^3}{24} X[K].
  \end{align}
\end{subequations}
It is straightforward to transform between these two conventions via \cite{Hansen-GoosJPCM2006,OettelEL2009}
\begin{subequations}\label{eq:exclusion-transform}
  \begin{align}
    a_0'
    &=
    a_0
    - \frac{\sigma}{2} a_1
    + \frac{\sigma^2}{4} a_2
    - \frac{\sigma^3}{24} p,
    \\
    a_1'
    &=
    a_1
    - \sigma a_2
    + \frac{\sigma^2}{4} p,
    \\
    a_2'
    &=
    a_2 - \frac{\sigma}{2} p.
  \end{align}
\end{subequations}
The resulting $\Delta \Omega$ will be identical whichever surface is chosen, except when there is a topological change in the molecular surface marking the breakdown of the theory; this is discussed in detail in Ref.\ \cite{OettelEL2009}.

\subsection{Failure of previous morphometric theories in treating correlations}
\label{sec:spt-failure}

Having specified the surface, we can examine the self-consistency of correlation functions determined through previously known morphological theories.
We briefly state the main theories below, then proceed to show how they produce inaccurate correlation functions at high densities.
This underscores the need for a more accurate theory, and the specific inconsistency we highlight in this section will be used to construct one in the next section.

With either the scaled particle or morphometric \emph{ansatzes}, \eqref{eq:spt-ansatz} or \eqref{eq:morph-ansatz}, a specific theory comprises the set of coefficients $\{p,a_2,a_1,a_0\}$.
In appendix \ref{appendix:classical-spt} we summarise the classical scaled particle arguments of Refs.\ \cite{ReissJCP1959,LebowitzJCP1965} using modern notation, which produce coefficients
\begin{subequations}\label{eq:py-coefficients}
  \begin{align}
    \beta a_0^\mathrm{SPT/PY}
    &=
    -\frac{\ln{(1- \eta)}}{4\pi},
    \\
    \beta a_1^\mathrm{SPT/PY}
    &=
    \frac{3\eta}{2\pi \sigma (1 - \eta)},
    \\
    \beta a_2^\mathrm{SPT/PY}
    &=
    \frac{6\eta + 3\eta^2}{2\pi \sigma^2 (1 - \eta)^2},
    \\
    \frac{\beta p^\mathrm{SPT/PY}}{\rho}
    &=
    \frac{1 + \eta + \eta^2}{(1 - \eta)^3}.
    \label{eq:py-pressure}
 \end{align}
\end{subequations}
In this classical approach, the Percus-Yevick (PY) equation of state emerges as an \emph{output} of the theory.
More recently, morphometric theories have been obtained as the bulk limit of FMT, with the hitherto most successful theory determined in Ref.\ \cite{Hansen-GoosJPCM2006} as
\begin{subequations}\label{eq:cs-spt-coefficients}
  \begin{align}
    \beta a_0^\mathrm{SPT/CS}
    &=
    -\frac{\ln{(1- \eta)}}{4\pi},
    \\
    \beta a_1^\mathrm{SPT/CS}
    &=
    \frac{1}{2\pi\sigma} \left(
    \frac{5\eta + \eta^2}{1 - \eta}
    + 2 \ln{(1 - \eta)}
    \right),
    \\
    \beta a_2^\mathrm{SPT/CS}
    &=
    \frac{1}{\pi \sigma^2} \left(
    \frac{\eta (2 + 3\eta - 2\eta^2)}{(1 - \eta)^2}
    - \ln{(1 - \eta)}
    \right),
    \\
    \label{eq:cs-pressure}
    \frac{\beta p^\mathrm{SPT/CS}}{\rho}
    &=
    \frac{1 + \eta + \eta^2 - \eta^3}{(1-\eta)^3},
 \end{align}
\end{subequations}
obtained from a functional constructed to impose the Carnahan-Starling (CS) equation of state \eqref{eq:cs-pressure}.
The latter equation of state is known to be highly accurate across the whole stable liquid regime, and even at the high density limits accessible to simulation in the supercooled regime \cite{BerthierPRL2016}.
The same equations are also obtained in the bulk limit of the functional of Ref.\ \cite{SantosPRE2012}, which similarly imposes the CS pressure but is slightly more self-consistent.
Curiously, we can make a minor modification to SPT arguments to impose the CS equation of state as an \emph{input} to obtain the above coefficients without invoking FMT (details in appendix \ref{appendix:cs-spt}).
We thus label this theory as SPT/CS.

To demonstrate the inaccuracy of the correlation functions produced by these known theories using \eqref{eq:distribution-functions}, we consider what happens to the pair correlation at high densities.
The potential of mean force \eqref{eq:potential-mean-force} for non-overlapping spheres with the morphometric \emph{ansatz} \eqref{eq:morph-ansatz} is written
\begin{equation}\label{eq:phi2}
  \begin{split}
    \phi^{(2)}(r)
    :=&
    - k_B T \ln g^{(2)}(r)
    \\ =&
    p V(r) + a_2 A (r) + a_1 C(r) + a_0 X(r) - 2\mu^\mathrm{ex}[p].
  \end{split}
\end{equation}
As a self-consistency test, we will compare this explicit result at contact against the \emph{exact} value of $g^{(2)}(\sigma)$ predicted by the \emph{virial theorem} as \cite{Hansen2013}
\begin{equation}\label{eq:contact-g}
  g^{(2)}(\sigma) =
  \frac{3}{2\pi \sigma^3 \rho} \left( \frac{\beta p}{\rho} - 1 \right).
\end{equation}

To evaluate \eqref{eq:phi2} we need to calculate the size measures for the two particle solute resembling a ``dumbbell''.
It is easier to calculate the excluded volume geometry, after which we can obtain the molecular volumes using the canonical relations \eqref{eq:exclusion-transform-volumes}.
The excluded volume consists of the union of two balls of radius $\sigma$ separated by a distance $r$.
The geometric properties at contact are then \cite{OettelEL2009}
\begin{align*}
  X_+(\sigma)
  &=
  4 \pi
  \\
  C_+(\sigma)
  &=
  \left( 6 - \frac{\pi}{2\sqrt{3}} \right) \pi \sigma,
  \\
  A_+(\sigma)
  &=
  6 \pi \sigma^2,
  \\
  V_+(\sigma)
  &=
  \frac{9 \pi \sigma^3}{4}.
\end{align*}
Transforming to the parallel molecular surface using the inverse transformation of \eqref{eq:exclusion-transform-volumes} gives the solute parameters as
\begin{subequations}\label{eq:contact-volumes}
  \begin{align}
    C(\sigma)
    &=
    \left( 4 - \frac{\pi}{2\sqrt{3}} \right) \pi \sigma,
    \\
    A(\sigma)
    &=
    \left( 1 + \frac{\pi}{2\sqrt{3}} \right) \pi \sigma^2,
    \\
    V(\sigma)
    &=
    \left( \frac{7}{12} - \frac{\pi}{8\sqrt{3}} \right) \pi \sigma^3.
  \end{align}
\end{subequations}
Fig.\ \ref{fig:contact-g} shows the contact value $g^{(2)}(\sigma)$ from inserting the geometric parameters above into \eqref{eq:phi2}, and the quasi-exact result of \eqref{eq:contact-g} assuming the CS equation of state \eqref{eq:cs-pressure}.
We find that both the known morphometric theories \eqref{eq:py-coefficients} and \eqref{eq:cs-spt-coefficients} are reasonably accurate until around the freezing density, above which contact correlations spuriously decay.
Thus, a new theory is needed to treat correlations at high densities; in the next section we will construct one which satisfies \eqref{eq:contact-g} by construction.

\begin{figure}
  \includegraphics[width=\linewidth]{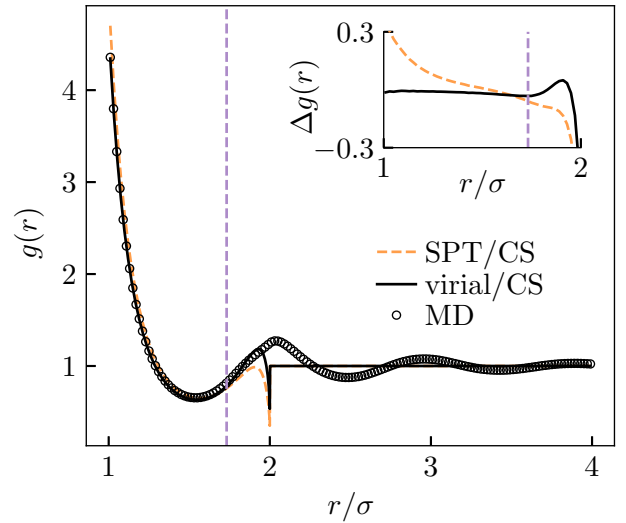}
  \caption{Comparing radial distribution functions of the morphometric theories  which impose the Carnahan-Starling equation of state \eqref{eq:cs-pressure}, against results of molecular dynamics (MD) simulations at volume fraction $\eta = 0.45$.
    The inset shows the difference between the two theoretical distribution functions and the molecular dynamics.
    The purple dashed line indicates where the molecular surface self-intersects at $r = \sqrt{3} \sigma$, marking the end of the theory's regime of validity.}
  \label{fig:g2-full}
\end{figure}

\subsection{Obtaining the new theory by self-consistency of the contact value of $g^{(2)}(r)$ with the virial theorem}
\label{sec:virial-spt}

Our goal is to develop a morphometric theory which produces accurate correlation functions $g^{(n)}$.
As described at the end of the last section, the correlation functions produced by an SPT approach are inaccurate at high densities.
We will correct the spurious decay of the contact value of the pair distribution function $g^{(2)}(r)$ at high densities by building this into the theory explicitly, with the aim of producing more accurate correlation functions.
A working understanding of scaled particle arguments is necessary to follow the details of this derivation, which we lay out in appendices \ref{appendix:classical-spt} and \ref{appendix:cs-spt}.

Inserting the volumes at contact \eqref{eq:contact-volumes} into \eqref{eq:phi2} and applying the virial theorem \eqref{eq:contact-g} for the contact value of $g^{(2)}$ gives the final expression
\begin{widetext}
  \begin{equation}\label{eq:v-virial}
  p \left( \frac{7}{12} - \frac{\pi}{8\sqrt{3}} \right) \pi \sigma^3
  + a_2 \left( 1 + \frac{\pi}{2\sqrt{3}} \right) \pi \sigma^2
  + a_1 \left( 4 - \frac{\pi}{2\sqrt{3}} \right) \pi \sigma
  + a_0 \, 4\pi
  =
  2\mu^\mathrm{ex}[p] - \beta^{-1} \ln{\frac{3}{2\pi \rho \sigma^3} \left( \frac{\beta p}{\rho} - 1 \right)}.
\end{equation}
We will use this last expression instead of the contact theorem \eqref{eq:spt-virial} in order to obtain new coefficients.
Together \eqref{eq:spt-point}, \eqref{eq:spt-mu} and \eqref{eq:v-virial} solve to give coefficients:
\begin{subequations}\label{eq:virial-coefficients}
  \begin{align}
    \beta a_0^\mathrm{virial}
    &=
    -\frac{\ln{(1- \eta)}}{4\pi},
    \\
    \beta a_1^\mathrm{virial}
    &=
    \frac{1}{(\sqrt{3}\pi - 4)\pi\sigma}
    \left(
    \left(5 - \frac{5\pi}{2\sqrt{3}}\right)
    \eta \frac{\beta p}{\rho}
    - \left( 2 - \frac{\pi}{\sqrt{3}} \right) \beta \mu^\mathrm{ex}[p]
    + \frac{\pi}{\sqrt{3}} \ln{(1 - \eta)}
    + 2 \ln{\left( \frac{\frac{\beta p}{\rho} - 1 }{4\eta} \right)}
    \right),
    \\
    \beta a_2^\mathrm{virial}
    &=
    - \frac{1}{(\sqrt{3}\pi - 4)\pi \sigma^2}
    \left(
    \left(6 - \frac{2\pi}{\sqrt{3}}\right)
    \eta \frac{\beta p}{\rho}
    - \frac{\pi}{\sqrt{3}} \beta \mu^\mathrm{ex}[p]
    + \left( 4 - \frac{\pi}{\sqrt{3}} \right) \ln{(1 - \eta)}
    + 4 \ln{\left( \frac{\frac{\beta p}{\rho} - 1 }{4\eta} \right)}
    \right).
 \end{align}
\end{subequations}
\end{widetext}
We refer to coefficients obtained this way for the CS pressure \eqref{eq:cs-pressure} as virial/CS, but we will not give them explicitly.
Unlike the WBII coefficients above these are new.
The pair correlation produced by these coefficients (black line in Fig.\ \ref{fig:contact-g}) is self-consistent with CS at contact by construction.

\section{Numerical results}
\label{sec:numerics}

\begin{figure}
  \includegraphics[width=\linewidth]{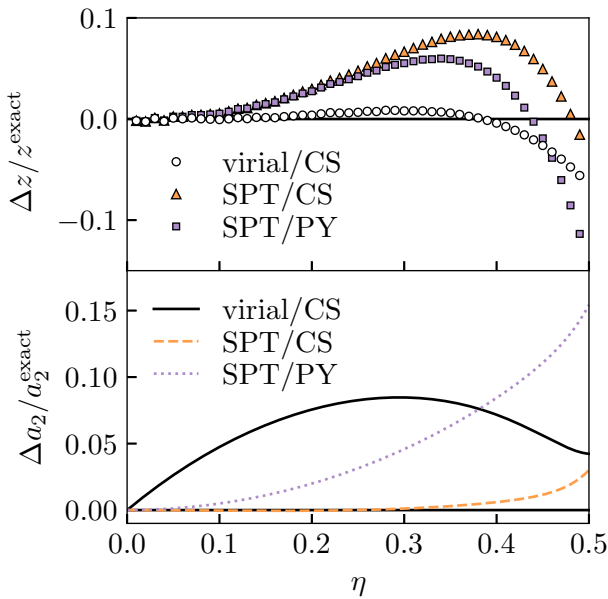}
  \caption{Errors in different morphometric theories for hard spheres.
    Top panel: error in the coordination defined in \eqref{eq:coordination}, giving the average number of neighbours in the shell $r < 1.4\sigma$ around a particle.
    Bottom panel: planar surface tensions against volume fraction, using the highly accurate result \eqref{eq:quasi-exact-surface-tension} from Ref.\ \cite{DavidchackMP2015} valid until $\eta \sim 0.5$.
  }
  \label{fig:surface-tension}
\end{figure}

We apply the thermodynamic coefficients determined in previous sections for a system of hard spheres to obtain two-- and three--body distribution functions using the generalised potential of mean force \eqref{eq:potential-mean-force} with the morphometric approach \eqref{eq:morph-ansatz}, and compare these against molecular dynamics simulations.
For the analytics we determine the input geometric quantities $\{V,A,C,X\}$ using the algorithms of Refs.\ \cite{MeckeAA1994,KleninJCC2011}.
For the simulations we performed event-driven molecular dynamics of $N=1372$ monodisperse hard spheres using the DynamO software package \cite{BannermanJCP2010}.
We measure the pair and triplet distribution functions $g^{(2)}$ and $g^{(3)}$ for simulations at $\eta = 0.45$.
For simulations above freezing $\eta \simeq 0.494$ we used a 5-component equimolar distribution with $\sim8\%$ polydispersity.

For $g^{(2)}$ shown in Fig.\ \ref{fig:g2-full} we find the virial/CS theory outperforms the SPT/CS theory even away from contact.
The agreement with the molecular dynamics simulations is excellent, until $r \gtrsim \sqrt{3}\sigma$ where the solute boundary self-intersects marking the end of the theory's regime of validity.
Geometrically, the regime $r < \sqrt{3}\sigma$ is the regime where the canonical relations \eqref{eq:exclusion-transform-volumes} apply so the thermodynamics is independent of the choice of surface definition.
Physically, for $r > \sqrt{3}\sigma$ interactions between solvent particles can occur \emph{through} the solute, and these correlations are not captured by the theory.
More discussion of this breakdown can be found in Ref.\ \cite{OettelEL2009}.
Only the contact value was fixed, so accuracy for $r > \sigma$ was not guaranteed; the accuracy is a welcome bonus.
We can quantify this accuracy through the integrated value
\begin{equation}\label{eq:coordination}
  z(\delta)
  =
  4\pi \int_\sigma^{\sigma+\delta}
  \rho^{(2)}(r) \, r^2 \, dr
\end{equation}
shown in the top panel of Fig.\ \ref{fig:surface-tension} where we take $\delta = 0.4\sigma$.
We find this integrated quantity is within $10\%$ accuracy across the liquid regime for all three theories, with the new theory performing substantially better overall.
Despite the improved accuracy, the errors begin to increase in magnitude at the end of the liquid regime so we expect them to become significant with very deep supercooling.

Next we compare the theories' predicted surface tension against simulation data.
The surface tension at a planar wall is simply $a_2$ because it conjugates with the area.
In Ref.\ \cite{DavidchackMP2015} a highly accurate $a_2$ was measured for hard spheres through extensive simulation, which was parameterised by the following expression
\begin{widetext}
\begin{equation}\label{eq:quasi-exact-surface-tension}
  \beta a_2
  =
  \frac{1}{\pi \sigma^2} \left(
  \frac{\eta (2 + 3\eta - \frac{9}{5}\eta^2 - \frac{4}{5}\eta^3 - (5 \times 10^4) \eta^{20})}{(1 - \eta)^2}
  - \ln{(1 - \eta)}
  \right).
\end{equation}
\end{widetext}
Comparing this highly accurate expression against the values predicted from the morphometric coefficients, we find the virial/CS surface tension is less accurate than the SPT/CS prediction (Fig.\ \ref{fig:surface-tension} bottom panel) despite its superior correlation functions at high densities.
Moreover, we find that at low densities the new theory is less accurate than classical SPT/PY theory.
This discrepancy occurs because both SPT/PY and SPT/CS feature the correct low density asymptotics of $a_2 \sim \mathcal{O}(\eta)$, which is imposed through the radial derivative of $\Delta\Omega(r)$ in the point solute limit \eqref{eq:spt-point-d1}.
This suggests that the new virial/CS theory sacrifices asymptotic accuracy at low densities, for more self-consistency of the surface tension at moderate to high densities.
One of the great strengths of the SPT/CS theory is its accuracy in the planar limit \cite{Hansen-GoosJPCM2006}, and so SPT/CS coefficients may give more accurate grand potentials (and thus correlations) for large solutes where the surface becomes approximately planar.

\begin{figure}
  \includegraphics[width=\linewidth]{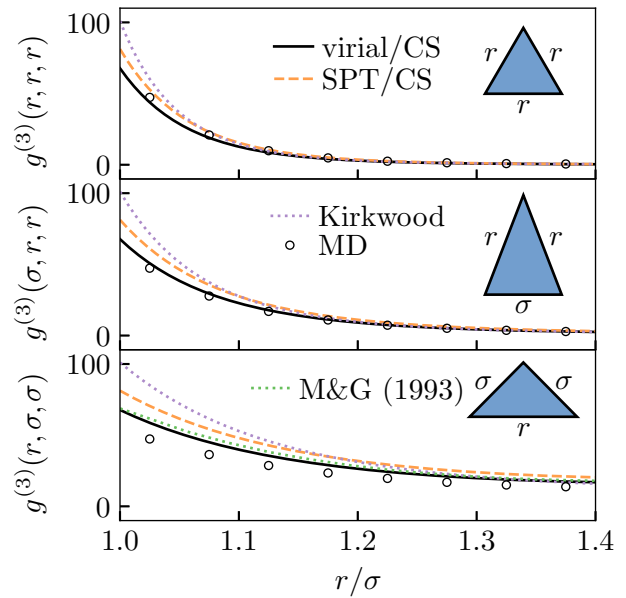}
  \caption{Comparison of predicted correlations for the morphometric approaches in triangular geometries, i.e.\ the first correlations beyond the pair level, against molecular dynamics simulations at volume fraction $\eta = 0.45$.
  In the bottom panel we also include the tabulated values of Ref.\ \cite{MullerMP1993} for comparison.}
  \label{fig:g3}
\end{figure}

Our goal was to develop a theory capable of treating correlations at the many-body level, so we now examine three-body correlation functions.
Triplet geometries are characterised by a triangle of side lengths $r,s,t$ so $g^{(3)} = g^{(3)}(r,s,t)$.
We also compare the morphometric theories against the Kirkwood approximation \cite{KirkwoodJCP1935} i.e.\
\begin{equation}\label{eq:kirkwood}
  g^{(3)}(r, s, t)
  \approx
  g^{(2)}(r) g^{(2)}(s) g^{(2)}(t)
\end{equation}
where we take the values of $g^{(2)}$ from the virial/CS theory because of its already demonstrated accuracy at the two body level.
Comparison of the morphometric correlation functions, and the Kirkwood closure, against molecular dynamics are shown in Fig.\ \ref{fig:g3}.
The virial/CS closure most closely matches the simulations at high densities, suggesting the theory is suitable for modeling complex many-particle local structures \cite{RobinsonPRL2019}.
For comparison we also include the tabulated values of Ref.\ \cite{MullerMP1993} where $g^{(3)}$ is used to treat polyatomic molecules \nocite{AttardCPL1992}
\footnote{We believe there is a misprint in Eq.\ (A1) of Ref.\ \cite{MullerMP1993}, which should read $(1-\eta)^3$ in the denominator.
  We refitted their source data \cite{AttardCPL1992} with this corrected form to be sure.}; our theory is marginally more accurate, and more importantly it provides a recipe for treating the higher-order correlation functions.

\begin{figure}
  \includegraphics[width=\linewidth]{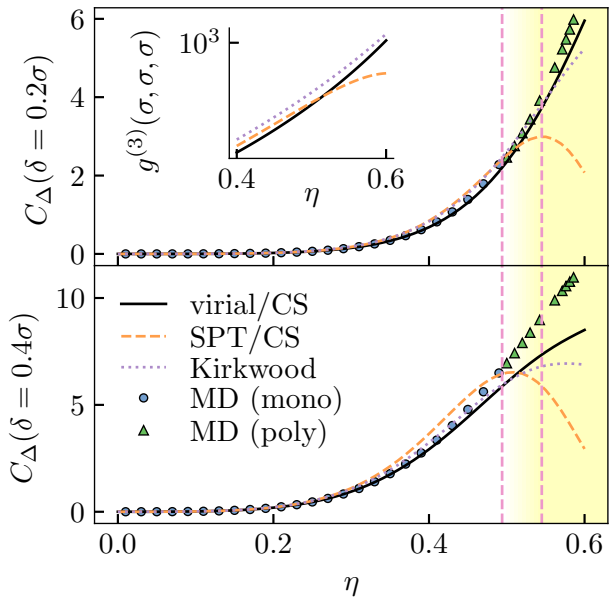}
  \caption{Concentration of triangles in the hard sphere liquid with side lengths $r,s,t \in [\sigma, \sigma + \delta]$ versus volume fraction.
    Direct measurements by molecular dynamics using a single-component system and an 8\% polydisperse system, while the lines show predictions from the morphometric theories described in text.
    The hard sphere freezing and melting volume fractions are indicated by pink dashed lines to show the onset of the supercooled regime.
    Inset: contact value of $g^{(3)}$ showing how the errors in the SPT/CS theory arise from underestimation close to contact.}
  \label{fig:sp3}
\end{figure}

To quantify accuracy at the three-body level we consider the concentration of triangles with side lengths $r,s,t \in [\sigma, \sigma + \delta]$ in the bulk liquid, from \eqref{eq:n-density-pdf} we find this as \cite{KrumhanslJCP1972}
\begin{equation}
  C_\Delta(\delta)
  =
  8\pi^2 \int_\sigma^{\sigma+\delta}\int_\sigma^{\sigma+\delta}\int_\sigma^{\sigma+\delta}
  \rho^{(3)}(r,s,t) \, rst \, dr ds dt.
\end{equation}
Comparison with molecular dynamics simulations in Fig.\ \ref{fig:sp3} shows similar levels of accuracy for small $\delta$, though the performance decreases as it is increased above the first minimum of the $g^{(2)}(r)$; this is not surprising as our virial closure only enforces accuracy approaching contact.
Notably, the Kirkwood approximation \eqref{eq:kirkwood} performs surprisingly well at the three-body level in both of these tests.

\section{Discussion and summary}

We have presented the morphometric approach as a generalisation of SPT, thus placing the scaled particle \emph{ansatz} on more precise and physically motivated assumptions i.e.\ those underlying the theorems of integral geometry.
Using the scaled particle approach we have systematically derived a new theory capable of accurately calculating many-body correlations in the hard sphere liquid; we recently used this to accurately treat local structures in Ref.\ \cite{RobinsonPRL2019}.
Our scaled particle formalism is flexible enough to derive all known morphometric theories without invoking fundamental measure theory.

In principle this approach could be extended to simple liquids where the interaction potential can be approximated as a perturbation around a hard core e.g.\ the Lennard-Jones potential.
However, as we exploited features of the hard sphere interaction potential to achieve closed form expressions for the thermodynamic coefficients, more realistic interaction potentials would likely require numerical expressions.
Additionally, attractions can introduce non-analytic behaviour from wetting/drying transitions which would not be accounted for in our theory \cite{EvansELE2003,EvansJCP2004}.

By making the underlying assumptions explicit we can better understand the limits of the theory: any deviation from the morphometric/SPT \emph{ansatz} must be due to a violation of translation/rotation invariance, additivity or continuity.
The fact that these theories are very accurate for hard spheres suggests that the assumptions are only weakly violated for this system.
While translational/rotational invariance and continuity are physically plausible conditions on $\Delta \Omega$, additivity is a very strong assumption.
In particular, we expect significant deviations from additivity where the liquid develops a static length scale exceeding the size of the solute \cite{KonigPRL2004}.
As such, we expect the validity of the morphometric approach to require the solute to be larger than the point-to-set length \cite{MontanariJSP2006}, which acts as an upper bound for all structural length scales \cite{YaidaPRE2016}.
The morphometric \emph{ansatz} must break down approaching a critical point, so it cannot be used to obtain asymptotics in the event of a thermodynamic glass transition.

Finally, we remark that while it is tempting to call the treatment of bulk degrees of freedom with the morphometric approach mean-field, this is not a completely accurate characterisation.
Mean-field theories typically become formally exact in the limit of infinite spatial dimensions, where the thermodynamic role of fluctuations disappears.
By contrast, the morphometric approach (and related theories like SPT and FMT) become formally exact in the one-dimensional limit of hard rods.
Though this theory does not explicitly describe fluctuations, they are built into the choice of thermodynamic coefficients entering the theory.
In this sense it is more accurate to describe the morphometric approach (and related theories) as an \emph{excluded volume} theory, or as a \emph{free volume} theory because the thermodynamics only shows divergent behaviour as $\eta \to 1$.

\appendix

\section{Justification of the scaled particle ansatz}
\label{appendix:spt-ansatz}

We now give a brief justification of our main approximation \eqref{eq:morph-ansatz}, in particular why there are only four terms in the expansion.
Radius is the only natural parameter for a sphere, however for more general geometries there might be arbitrarily many parameters so one may wonder if they should be included in a general geometric expansion.
Nevertheless, there are compelling arguments from integral geometry \cite{KonigPRL2004} to only retain the four terms listed which we will summarise below.

The basis of the morphometric approach is that the functionals $\{V,A,C,X\}$ are normalisations of the so-called \emph{intrinsic volumes}.
These play a central role in integral geometry as the \emph{only} physically meaningful size measures in the sense that they:
\begin{enumerate}
\item Are invariant with respect to translations and rotations.
\item Increase additively, i.e.\ they transform under combination of subsystems via the inclusion/exclusion relation e.g.\
  \begin{equation*}\label{eq:additivity}
    V[A \cup B] = V[A] + V[B] - V[A \cap B],
  \end{equation*}
  and similar expressions for $A$, $C$, and $X$.
\item Are continuous (specifically with respect to the Hausdorff metric).
  Loosely speaking, this means that the size measures converge as the object is approximated by increasingly finely meshed polyhedra excluding e.g.\ fractal geometries.
  As a simple intuitive example, the measurement of a length will converge continuously to some number as one uses rulers with progressively finer distance markings.
\end{enumerate}
More details on properties of intrinsic volumes can be found in standard texts, e.g.\ Refs.\ \cite{Santalo2004,Klain1997}.

The central assertion of the morphometric approach is that the insertion cost $\Delta \Omega$ \emph{exactly} possesses the properties above, providing the connection between geometry and thermodynamics \cite{KonigPRL2004}.
A classic theorem of integral geometry due to Hadwiger \cite{Hadwiger1957} states that the intrinsic volumes are the \emph{only} class of functionals with the properties listed above; a corollary of this is that they form a linear vector space for any functional possessing these properties.
The morphometric form \eqref{eq:morph-ansatz} then follows.
In addition to providing a more general \emph{ansatz} than SPT, this approach lays out its underlying assumptions explicitly eschewing the ad-hoc way in which the original SPT \emph{ansatz} \eqref{eq:spt-ansatz} was obtained.
Moreover, classical SPT assumes hard spheres from the outset while our generalisation based on integral geometry is more flexible, allowing for generalisations to mixtures, more realistic pair potentials and non-spherical particles without compromising its assumptions.

The morphometric approach is certainly an approximation, as the insertion cost will not rigorously possess the three properties above in reality.
Notably, in SPT $\Delta\Omega$ is known to contain singularities in its high order derivatives with solute radius \cite{ReissJCP1959}; these non-analytic terms result from violations of the additivity assumption.
Nevertheless, the approximation is accurate in hard spheres \cite{OettelEL2009,AshtonPRE2011,LairdPRE2012,BlokhuisPRE2013,UrrutiaPRE2014,Hansen-GoosJCP2014} so these violations should be small.

\section{Classical scaled particle relations}
\label{appendix:classical-spt}

Following the protocol of scaled particle theories, we consider the insertion of a hard spherical solute of radius $R$ into the liquid.
Assuming the morphometric form for the insertion cost returns us to the \emph{ansatz} \eqref{eq:spt-ansatz}.
Below we give the exact thermodynamic relations for hard spheres which produce the classical SPT coefficients.

It is possible to consider the insertion of a solute with a \emph{negative} radius: the hard core interaction between the two particles only occurs when the solute is `inside' a solvent particle.
In this limit the insertion cost can be determined exactly as \cite{ReissJCP1959}
\begin{equation}
  \beta \Delta \Omega =
  -\ln{\left(
    1 - \frac{4\pi \left(R + \frac{\sigma}{2}\right)^3}{3} \rho
    \right)}
\end{equation}
for $-\frac{\sigma}{2} \le R \le 0$.
It may appear concerning that this result does not possess the morphometric form \eqref{eq:morph-ansatz}; however, this does not discount the validity of the morphometric approach as the nonphysical geometry violates the continuity assumption (section~\ref{sec:morphometric-approach}) because it cannot be approximated by polyhedra.
This places the result for $R < 0$ outside the theory's stated regime of validity, however $\Delta\Omega$ is continuous up to its second derivative across $R=0$ with a discontinuity in its third derivative \cite{ReissJCP1959}.
In the limit $R \to 0$ the expression above corresponds to the cost of inserting a hard point giving
\begin{subequations}\label{eq:spt-origin-continuity}
  \begin{align}
    \label{eq:spt-point}
    \beta \Delta\Omega(R=0)
    &=
    -\ln{(1- \eta)},
    \\
    \label{eq:spt-point-d1}
    \beta \left. \left(
    \frac{\partial \Delta\Omega}{\partial R}
    \right)_{\mu, V, T} \right|_{R=0}
    &=
    \frac{6\eta}{\sigma (1- \eta)},
    \\
    \label{eq:spt-point-d2}
    \beta \left. \left(
    \frac{\partial^2 \Delta\Omega}{\partial R^2}
    \right)_{\mu, V, T} \right|_{R=0}
    &=
    \frac{12\eta^2 + 24\eta}{\sigma^2 (1- \eta)^2}.
 \end{align}
\end{subequations}
Note that \eqref{eq:spt-point} can also be justified by considering that the probability of a randomly selected position in space being empty is simply the free volume $1-\eta$.

Together applying \eqref{eq:spt-origin-continuity} to \eqref{eq:spt-ansatz} fixes the coefficients $\{a_0, a_1, a_2\}$, so the theory requires an additional thermodynamic relation to determine the pressure.
When $R = \frac{\sigma}{2}$ the solute is equivalent to the solvent particles themselves and we recover $\Delta\Omega = \mu^\mathrm{ex}$, so from \eqref{eq:spt-ansatz} we have
\begin{equation}\label{eq:spt-mu}
  \Delta\Omega\left(R=\frac{\sigma}{2}\right) =
  \frac{\pi \sigma^3}{6} p
  + \pi \sigma^2 \, a_2
  + 2 \pi \sigma \, a_1
  + 4\pi \, a_0
  =
  \mu^\mathrm{ex}.
\end{equation}
Combining this expression with the thermodynamic relation \eqref{eq:osmotic-consistency-2} gives a differential equation for $\beta p$ whose solution gives the classical SPT coefficients for hard spheres \eqref{eq:py-coefficients}.
The equation of state \eqref{eq:py-pressure} is equivalent to the one obtained through the solution of the Percus-Yevick (PY) integral equation \cite{WertheimPRL1963}; these two routes have been unified within FMT \cite{RosenfeldPRL1989}.

\section{First generalisation: SPT with an empirical equation of state}
\label{appendix:cs-spt}

In the classical SPT approach described in the previous section, the SPT/PY equation of state emerges as an \emph{output} of the theory.
Taking inspiration from the White Bear free energy functional \cite{RothJPCM2002}, we reformulate the SPT argument so that the equation of state is an \emph{input} to the theory.
In so doing we aim to construct a theory from a more accurate equation of state, with the trade-off being that we must sacrifice some self-consistency.
The main equation of state we impose is the CS relation \eqref{eq:cs-pressure} \cite{CarnahanJCP1969}.
This ultimately results in a theory previously known as a limit of a free energy functional \cite{Hansen-GoosJPCM2006}, but through simpler arguments.
We extend these arguments in the main text to arrive at a new theory capable of accurately treating correlation functions.

A crucial component of scaled particle approaches is thermodynamic consistency of the (osmotic) pressure via
\begin{equation}\label{eq:osmotic-consistency-1}
  \beta p
  =
  \rho - \beta f^\mathrm{ex}
  + \rho \left( \frac{\partial \beta f^\mathrm{ex}}{\partial \rho} \right)_{V,T}
\end{equation}
where $f^\mathrm{ex} = F^\mathrm{ex}/V$ is the (excess) free energy density.
The form most useful for a single-component system comes from taking the derivative with respect to density, and noting the definition of the excess chemical potential
\begin{equation*}
  \beta \mu^\mathrm{ex}
  =
  \left( \frac{\beta f^\mathrm{ex}}{\partial \rho} \right)_{V,T},
\end{equation*}
giving
\begin{equation}\label{eq:osmotic-consistency-2}
    \left( \frac{\partial \beta p}{\partial \rho} \right)_{V,T}
    =
    \rho \left( \frac{\partial \beta \mu}{\partial \rho} \right)_{V,T}
    =
    1 + \rho \left( \frac{\partial \beta \mu^\mathrm{ex}}{\partial \rho} \right)_{V,T}.
\end{equation}
Note that the consistency relation \eqref{eq:osmotic-consistency-1} provides the route to generalising all of our arguments to arbitrary mixtures \cite{RosenfeldPRL1989,SollichAiCP2001,SantosPRE2012}.
The free energy remains well-defined, even for polydisperse mixtures, if the composition dependence enters only through a set of weighted moments of the density $\{\xi_k\}$.
Then \eqref{eq:osmotic-consistency-1} becomes
\begin{equation}\label{eq:osmotic-consistency-3}
  \beta p
  =
  \rho - \beta f^\mathrm{ex}
  + \sum_k
  \xi_k \left( \frac{\partial \beta f^\mathrm{ex}}{\partial \xi_k} \right)_{V,T}.
\end{equation}
For weighted densities consistent with an SPT (or FMT) approach, it has been shown Ref.\ \cite{SantosPRE2012} that the thermodynamic coefficients for mixtures are determined once an equation of state for the single-component system is known.

We can relate the radial derivative of $\Delta\Omega(R)$ to the solvent density at contact; by connecting this to the virial theorem we can obtain a new thermodynamic relation.
Following Ref.\ \cite{BrykPRE2003} we take the normal derivative of $\Omega$ with respect to $R$, and noting that $\Delta\Omega(R) = \Omega(R) - \Omega_\mathrm{hom}$ gives
\begin{widetext}
\begin{equation}\label{eq:spt-derivative}
  \left( \frac{\partial \Delta \Omega}{\partial R} \right)_{\mu,V,T} =
  \left( \frac{\partial \Omega}{\partial R} \right)_{\mu,V,T} =
  \int
  \frac{\delta \Omega[\rho_0(\vec{r})]}{\delta \rho}
  \left( \frac{\partial \rho_0(\vec{r})}{\partial R} \right)_{\mu,V,T}
  \, d\vec{r} +
  \int
  \rho_0(\vec{r})
  \left( \frac{\partial \phi_\mathrm{ext}(\vec{r}; R)}{\partial R} \right)_{\mu,V,T}
  \, d\vec{r},
\end{equation}
\end{widetext}
where $\rho_0$ is the equilibrium density profile and $\phi_\mathrm{ext}$ is the external potential (i.e.\ the potential of the solute).
In equilibrium $\Omega$ is minimised so
\begin{equation*}
  \left.
  \frac{\delta \Omega[\rho(\vec{r}); \phi_\mathrm{ext}]}{\delta \rho}
  \right|_{\rho(\vec{r})=\rho_0(\vec{r})} = 0,
\end{equation*}
and the first integral in \eqref{eq:spt-derivative} vanishes.
As the solute is hard, the external potential and its derivative are zero everywhere except at a distance $\frac{\sigma}{2}$ from the surface where both $\rho_0$ and $\phi_\mathrm{ext}$ are discontinuous.
We consider its Boltzmann weight, i.e.\
\begin{equation*}
  e^{-\beta\phi_\mathrm{ext}(\vec{r})}
  =
  \Theta\left( |\vec{r}| - R - \frac{\sigma}{2} \right).
\end{equation*}
Taking the (distributional) derivative of both sides gives
\begin{equation*}
  \beta\left( \frac{\partial\phi_\mathrm{ext}(\vec{r})}{\partial R} \right)_{\mu,V,T}
  =
  \delta\left( |\vec{r}| - R - \frac{\sigma}{2} \right)
  e^{\beta\phi_\mathrm{ext}(\vec{r})}.
\end{equation*}
Inserting this expression into \eqref{eq:spt-derivative} and using the fact that $\rho(\vec{r}) e^{\beta\phi_\mathrm{ext}(\vec{r})}$ is a continuous function (cf.\ Ref.\ \cite{Hansen2013}) gives the contact theorem
\begin{equation*}
  \beta \left( \frac{\partial \Omega}{\partial R} \right)_{\mu,V,T}
  =
  4\pi \left( R + \frac{\sigma}{2} \right)^2
  \rho\left( R + \frac{\sigma}{2} \right)
\end{equation*}
and the contact density in this inhomogeneous system is $\rho(\sigma; \phi_\mathrm{ext}) := \rho^{(2)}(\sigma)/\rho = \rho \, g^{(2)}(\sigma)$ recalling the definition of $\rho^{(n)}$ for the homogeneous system \eqref{eq:n-density-pdf} (cf.\ Ref.\ \cite{PercusPRL1962}), so we have
\begin{equation*}\label{eq:spt-contact-density}
  \begin{split}
    \left. \beta \left( \frac{\partial \Delta \Omega}{\partial R} \right)_{\mu,V,T}
    \right|_{R = \frac{\sigma}{2}}
    &=
    \left. \beta \left( \frac{\partial \Omega}{\partial R} \right)_{\mu,V,T}
    \right|_{R = \frac{\sigma}{2}}
    \\ &=
    4\pi \sigma^2 \rho \, g^{(2)}(\sigma).
  \end{split}
\end{equation*}
So inserting the SPT \emph{ansatz} \eqref{eq:spt-ansatz} gives
\begin{equation}\label{eq:spt-virial-g}
  \pi \sigma^2 \, p
  + 4\pi \sigma \, a_2
  + 4\pi \, a_1
  =
  \frac{4\pi \sigma^2 \rho}{\beta} \, g^{(2)}(\sigma).
\end{equation}
Inserting the virial theorem \eqref{eq:contact-g} into the right-hand side of \eqref{eq:spt-virial-g} yields the final expression:
\begin{equation}\label{eq:spt-virial}
  \pi \sigma^2 \, p
  + 4\pi \sigma \, a_2
  + 4\pi \, a_1
  =
  \frac{6}{\beta\sigma} \left( \frac{\beta p}{\rho} - 1 \right).
\end{equation}
This relation is satisfied by the coefficients \eqref{eq:py-coefficients}, which is surprising given that it was obtained from a completely different thermodynamic route and the \emph{ansatz} \eqref{eq:spt-ansatz} is inexact.
Nonetheless, this self-consistency is a testament to the effectiveness of SPT and related approaches.

Since the pressure is now a known input, the excess chemical potential can be determined by integrating \eqref{eq:osmotic-consistency-2} i.e.\
\begin{equation}\label{eq:excess-chemical-potential}
  \beta \mu^\mathrm{ex}[p]
  = \left( \frac{\beta p}{\rho} - 1 \right)
  + \int_0^\eta \left( \frac{\beta p}{\rho} - 1 \right) \, \frac{d\eta'}{\eta'}.
\end{equation}
To keep the expressions simple we will not evaluate the chemical potential until the very end, but it should be recognised as a known variable wherever it appears.
With the pressure fixed we have three free parameters in the theory $\{a_0, a_1, a_2\}$;
we must thus choose three out of the five available thermodynamic relations in \eqref{eq:spt-origin-continuity}, \eqref{eq:spt-mu} and \eqref{eq:spt-virial} to satisfy.
Therefore, we must lose consistency with two of these relations to obtain a more accurate theory for practical applications.

To set the correct energy scale we choose to fix $\Delta \Omega(R=0)$ and $\Delta \Omega(R=\sigma/2)$ through equations \eqref{eq:spt-point} and \eqref{eq:spt-mu} using the chemical potential determined above in \eqref{eq:excess-chemical-potential}.
This in turn imposes the consistency of the osmotic pressure \eqref{eq:osmotic-consistency-1}.
For the final equation we choose to set the contact value of $g^{(2)}$ through \eqref{eq:spt-virial} which better represents solutes of interest than the two relations for point geometries at $R=0$.
Solving these three equations gives the generalised SPT coefficients
\begin{subequations}\label{eq:general-spt-coefficients}
  \begin{align}
    \beta a_0^\mathrm{SPT}
    &=
    -\frac{\ln{(1- \eta)}}{4\pi},
    \\
    \beta a_1^\mathrm{SPT}
    &=
    \frac{1}{2\pi\sigma} \left(
    (\eta - 3) \frac{\beta p}{\rho}
    + 2 \beta \mu^\mathrm{ex}[p]
    + 2 \ln{(1 - \eta)}
    + 3
    \right),
    \\
    \beta a_2^\mathrm{SPT}
    &=
    - \frac{1}{\pi \sigma^2} \left(
    (2 \eta - 3) \frac{\beta p}{\rho}
    + \beta \mu^\mathrm{ex}[p]
    + \ln{(1 - \eta)}
    + 3
    \right).
 \end{align}
\end{subequations}
It can be verified that inserting the Percus-Yevick equation of state \eqref{eq:py-pressure} into these expressions yields the previously obtained coefficients \eqref{eq:py-coefficients}, as expected.
Inserting the CS equation of state we obtain \eqref{eq:cs-spt-coefficients} which are \emph{identical} to the coefficients derived from the White Bear II (WBII) free energy functional of Ref.\ \cite{Hansen-GoosJPCM2006}, although this is only clear after transforming to the excluded geometry through the canonical relations \eqref{eq:exclusion-transform}.
Remarkably, we have obtained these coefficients through a route completely different from their original derivation.

In Ref.\ \cite{Hansen-GoosJPCM2006} the coefficients were determined within FMT by taking the limit of a binary mixture where one component is infinitely dilute.
Here we completely avoided FMT, in favour of geometrical arguments similar to the classical SPT approach outlined in the previous section.
This suggests that this generalised scaled particle argument is built into the structure of the WBII functional of Ref.\ \cite{Hansen-GoosJPCM2006}; this is not an obvious fact as the derivation of this functional did not explicitly involve these arguments.
Rather, the WBII functional was constructed based on a novel extension of the CS equation to mixtures by requiring self-consistency of the pressure in \eqref{eq:osmotic-consistency-3} \cite{Hansen-GoosJCP2006}.
We imposed this relation by setting the chemical potential in \eqref{eq:spt-mu} using the chemical potential obtained via \eqref{eq:osmotic-consistency-2}.
It is unclear to us how our final choice of using \eqref{eq:spt-mu} instead of one of the two relations at the origin, i.e.\ \eqref{eq:spt-point-d1} or \eqref{eq:spt-point-d2}, is built into the WBII functional.

\begin{acknowledgements}
  We are grateful to Bob Evans who encouraged writing up the results of section \ref{sec:morphometric-approach}, which eventually became this manuscript.
  JFR, FT and CPR  acknowledge the European Research Council under the FP7 / ERC Grant Agreement No.\ 617266 ``NANOPRS''.
  CPR would like to acknowledge the Royal Society for financial support.
\end{acknowledgements}

\bibliography{bibliography}

\end{document}